\documentclass[12pt,preprint]{aastex}

\def\lsim{\hbox{ \rlap{\raise 0.425ex\hbox{$<$}}\lower 0.65ex\hbox{$\sim$} }}
\def\gsim{\hbox{ \rlap{\raise 0.425ex\hbox{$>$}}\lower 0.65ex\hbox{$\sim$} }}

\def\opeqn{\begin{equation}}
\def\cleqn{\end{equation}}

\begin{document}

\title{ Line Caustic Revisted: 
      What is $\delta$ in $J^{-1}\propto \sqrt{\delta^{-1}}$~?}
\author{Sun Hong Rhie (UND)}


\begin{abstract}
The line caustic behavior has been discussed since Chang and Refsdal (1979)
mentioned inverse-square-root-of-the-distance dependence of the amplification 
of the images near the critical curve in a study of a single point mass under  
the influence of a constant shear due to a larger mass. A quarter century 
later, Gaudi and Petters (2001) interprets that the distance is {\it a vertical 
distance to the caustic}. It is an erroneous misinterpretation.

We rehash Rhie and Bennett (1999) where the caustic behavior of the 
binary lenses was derived to study the feasibility of 
limb darkening measurements in caustic crossing microlensing events. 
~({\it 1}) $J = \pm \sqrt{4\delta\omega_{2-} J_-}$ where 
~$\delta\omega\parallel\bar\partial J$, and $\delta\omega_{2-}$ and $J_-$ are 
$E_-$-components of $\delta\omega$ (the source position shift from the caustic 
curve) and $2\bar\partial J$ (the gradient of the Jacobian determinant)
respectively;
~({\it 2}) The critical eigenvector $\pm E_-$ is normal to the caustic curve 
and easily determined from the analytic function $\kappa$-field; 
~({\it 3}) Near a cusp ($J_- = 0$) is of a behavior of the third order, and
the direction of $\bar\partial J$ with respect to the caustic curve changes 
rapidly because a cusp is an accumulation point;
~({\it 4}) On a planetary caustic, $|\partial J|\sim \sqrt{1/\epsilon_{pl}}$  
is large and power expansion does not necessarily converge over the size of
the lensed star. In practice, direct numerical summation is inevitable. 

We also note that a lens equation with constant shear is intrinsically   
incomplete and requires supplementary physical assumptions and interpretations
in order to be a viable model for a lensing system.      
\end{abstract}


\clearpage

\section*{ 0. \ \  {\it ``A Tale of Two Curves"} }

\begin{quotation}
\noindent
{\it There are two normals. One is the critical eigenvector and normal to
the caustic curve. The other is gradient J and normal to the critical curve.
The angle between the two normals changes along the curves.  
When the two normals are across, the caustic curve stops and turns around,
and the punctuation point is called a cusp.   
}
\end{quotation}

\section*{ -1. \ \  Back at Table with RB99}

The ingredients are a 2-d plane, two 2-d variables called source position 
and image position defined on the 2-d plane, a relation between the source 
position and image position called the lens equation, critical curve of the
lens equation where the lens equation is stationary (or {\it degenerate}), and 
caustic curve onto which the critical curve is mapped under the lens equation. 
The lens equation happens to be an explicit function from an image position
to its source position: $\omega = f(z, \bar z)$;~ 
$\bar\omega = \overline{f(z, \bar z)}$ \ where $z$ is an image position and
$\omega$ is its source position. The lens equation can be considered a mapping 
from the complex plane ($z$-plane because it is parameterized by $z$) to 
itself ($\omega$-plane because it is parameterized by $\omega$). This complex
plane (or the underlying 2-d real plane) is referred to as the lens 
plane because that is where all the tranverse objects related to lensing
are defined and studied. It is in fact convenient to put the lens plane at the
distance of the center of the mass of the lensing objects especially when the
focus of the lensing study is on the lensing objects such as microlensing planet
systems or compact objects.    

The issue is the local behavior of the lens equation in the neighborhood 
of the critical points (points on the critical curve) of quasi-analytic 
lens equations. A quasi-analytic lens equation is by defintion a lens
equation that is not analytic itself but whose derivatives are all determined 
by one analytic function called $\kappa$-field and its derivatives. 
The $n$-point lens equations including constant shear are quasi-analytic.  

The critical curve is one of the family of equipotential curves where the 
potential is given by the magnitude of $\kappa$ and the variable along the 
critical curve is the phase angle of $\kappa$. The phase angle also determines 
the direction vectors $\pm E_\pm$ (rb99.7).
The criticality defines the critical direction $\pm E_-$ (rb99.5), and the
caustic curve is everywhere normal to the critical direction. 
An arbitrary infinitesimal deviation $dz$ from the critical curve causes $\omega$ 
to shift only in the tangent to the caustic curve (rb99.8). 
The whole dimension of $E_-$ direction or the $E_-$ component 
of an arbitrary $dz$ is projected out by the lens equation, and that is what is 
referred to be {\it degenerate} and why the Jacobian determinant 
of the lens equation vanishes on the critical curve.  
In other words, 1) if we restrict to the linear 
order, an arbitrary deviation $d\omega=\delta\omega_{1+} E_+$ from a caustic point
has infinitely many solutions $dz = dz_+E_+ + dz_-E_-$ 
where $2dz_+ = \delta\omega_{1+}$ and $dz_-$ is indeterminate;  
2) thus, we need to include the next nonvanishing order terms
to see the proper behavior of the lens equation in the neighborhood 
of the critical points.  The second order terms have been written out 
in rb99.10 and rb99.13, or in equation (\ref{eqLomega}) below. The third order
terms are shown in equation (\ref{eqThird}). 

It is worth pointing out  
that the orthogonal decomposition coefficients are always real as should be
clear from equation (\ref{eqDecomp}). The differential structure is about the 
underlying linearly independent two dimensional space whether parameterized by
two real variables or by a complex variable and its complex conjugate, 
and thus whether one uses real 
basis vectors (implicitly 2-column vectors $(1,0)$ and $(0,1)$ defined at one
point on the caustic curve and the corresonding point on the critical curve) 
or $\pm E_\pm$ that are tangent and normal vector fields on the caustic curve, 
we inevitably deal with the same two degerees of freedom, namely the tangent 
component and normal component. One of the obvious advantages 
of maximally utilizing the quasi-analytic nature of the quasi-analytic lens 
equations by employing complex coordinate systems is that we have explicit 
expressions of the basis vector fields $E_\pm$. For example, we know how 
they change along the caustic curve parameterized by $\varphi$ half the 
phase angle of the $\kappa$-field: $\kappa = |\kappa| \exp{i2\varphi}$.  
\opeqn
 E_+ = e^{-i\varphi} \ , \quad E_- = i e^{-i\varphi}; \qquad  
 d E_+ = - E_- d\varphi \ , \quad d E_- = E_+ d\varphi \ .   
\cleqn  
(We note that the basis vector fields are not necessarily smooth everywhere.)   

We do see a potential problem in the emphasis to have used real variables 
in \citet{vertical} because they are putting the focus in a worng place. 
If it was a methodology to generate an undisclosed impression 
on the readers against our having consistently employed the complex plane, 
it perhaps is a wronful deed. If \citet{vertical} stated 
it because it is simply true that they worked with real coordinates, the fault 
lies in their failure to clear the smoke and address the relevant issues: 
\begin{enumerate}
\item
intrinsic understanding of the quadratic behavior of the lens equation; 
\item
correct formulation of $\Delta\omega_-$ (rb99.16 and equation(\ref{eqJdJtwo})) 
and its proper interpretation;  
\item
inconsistent truncation prescription in the first paragraph in SEF.p.189.  
\end{enumerate}
We will address the correct issues\footnote{Not all questions grammatically
correct are right questions.} in the following, of course using complex 
coordinates ($z$ for the image space and $\omega$ for the source space) 
and the analytic function $\kappa$-field. As corollaries, the followings
will be clear.
\begin{enumerate}
 \item 
The claim by \citet{vertical} that they reproduced SEF.6.17 is free of content. 
\item
The claim by \citet{vertical} that $\delta \equiv |\Delta\omega_-|$ is 
a ``vertical distance" is erroneous. 
\end{enumerate}
\citet{vertical} warns the readers, ``{\it there
are not many people who know that the distance is not the shortest distance
but the vertical distance."} The distance $\delta$ is neither the shortest
distance nor a vertical distance as is described in the abstract and 
illustrated in Fig. 1.  When $\delta$ can be approximated by the shortest 
distance, the ``vertical distance" converges to $\delta$ as well.  

\paragraph{\underline {\it Regurgitation of RB99:}} \ \
In the case of the binary lens in rb99.Fig.1 where the 
two masses are comparable and their separation is $\sim {\cal O}(1)$ 
(where $1$ refers to the Einstein ring radius of the total mass), the gradient
of the Jacobian determinant $|\nabla J| \sim {\cal O}(1)$. 
Equation rb99.13 reads as follows when $dz_+$ is reinstated.  
\opeqn
 \delta\omega_- = \delta\omega_{2-} 
 = {1\over 4} \left( J_- (dz_-^2 - dz_+^2) + 2 J_+ dz_+ dz_- \right)
\cleqn
If we assume the linear approximation for $dz_+$ as in rb99, the quadratic 
equation in $dz_-$ can be solved in terms of $dz_+$ and $\delta\omega_-$. 
\opeqn
dz_- = -{J_+\over J_-} dz_+ \pm \sqrt{{|\nabla J|^2 dz_+^2\over J_-^2}  
                          + {4 \delta\omega_{2-}\over J_-}}
\label{eqMinusOne}
\cleqn  
Since $dz_-$ is real, as we have emphasized that $E_\pm$ decomposition components 
are real variables above and in equation (\ref{eqDecomp}), $dz_-$ has 
solutions when the inside of the square root is non-negative (rb99.14). 
\opeqn
  \delta\omega_- \partial_- J + |\partial J|^2 dz_+^2 \ge 0
\label{eqZero}
\cleqn 
The degenerate solution for which the equality of equation 
(\ref{eqZero}) holds amounts to a shift along the critical curve.  
\opeqn
 dz_- = -{J_+\over J_-} dz_+ \quad \Rightarrow \qquad 
  J_+ dz_+ + J_- dz_-  = 0 
\cleqn    
If $z_\circ$ is a critical point, 
then $z_\circ^\prime \equiv z_\circ + dz_+ (E_+ - J_+ E_-/J_-)$ is also 
a critical point. 

We rewrite $z_\circ + dz$ as a vector deviated from $z_\circ^\prime$ 
such that $z_\circ + dz = z_\circ^\prime + dz^\prime$. Then, $dz^\prime$
is a vector in the direction of $\pm E_-(z_\circ)$.
\opeqn
 dz^\prime (z_\circ^\prime) 
    = \pm E_- (z_\circ) ~\sqrt{{|\nabla J|^2 dz_+^2\over J_-^2}
                          + {4\delta\omega_{2-}\over J_-}}
\cleqn  
If $\omega_\circ = \omega(z_\circ)$ is the caustic point corresponding
to the ``old" critical point $z_\circ$, then 
$\omega_\circ^\prime \equiv \omega_\circ + 2 dz_+ E_+(z_\circ)$ is a 
new caustic point because $E_+(z_\circ)$ is tangent to the caustic curve.
We can easily calculate the new caustic point $\omega_\circ^\prime$ up
to the second order. Let's start for clarity with the general second 
order expansion in $d\bar z$ in equation (\ref{eqDomega}).  
\opeqn
 \delta\omega_2 \equiv {1\over 2} \bar\partial\bar\kappa d\bar z^2
  = {1\over 2} (-\bar\partial J) (dz_+ - i dz_-)^2  \  
\cleqn 
The second equality is obtained by using 
$\bar\partial J = - \kappa \bar\partial\bar\kappa$, $E_\pm$-decomposition
$dz = dz_+ E_+ + dz_- E_-$, and miscellaneous relations such as $E_- = i E_+$
that can be found in rb99 and section 2. In order to get the expression for  
the caustic point, we apply the condition $0 = J_+ dz_+ + J_- dz_-$  
and get $\delta\omega$ parameterized by $dz_+$. 
\opeqn
\delta\omega = \delta\omega_+ E_+ + \delta\omega_- E_- 
    = \left(2 dz_+ + {1\over 4}{|\nabla J|^2\over J_-^2} J_+ dz_+^2 \right) E_+ 
       - {1\over 4} {|\nabla J|^2\over J_-^2} J_- dz_+^2 ~E_-
\cleqn
We rewrite the equations for $\delta\omega_+$ and $\delta\omega_-$ as follows
for later use.
\begin{eqnarray}
 0 &=& (\delta\omega_+ - 2 dz_+ ) J_- + J_+ \delta\omega_-    \\
 \delta\omega_- &=& - {1\over 4} {|\nabla J|^2\over J_-}  dz_+^2
\label{eqCpoint}
\end{eqnarray}  

It should be clear that the role of $dz_+$ is mainly to cause a migration of the
cautic point and critical point from $\omega_\circ$ and $z_\circ$ to new ones
$\omega_\circ^\prime$ and $z_\circ^\prime$ respectively. In order to see
the uncluttered picture of the local behavior of the lens equation near a
critical point, we set $dz_+= 0$.

\subsection*{ -1. \ \ In Details We Trust: Brutally Honest Interpretations }

If $dz_+ =0$, $dz = dz_- E_-$, and from equation (\ref{eqLomega}),
\begin{eqnarray*} 
\delta\omega_{1\pm} = 0 \ ; &\ &
\delta\omega_{2\pm}   =  {1\over 4} dz_-^2 J_\pm \   \\ 
    \Longleftrightarrow \qquad
 \delta\omega &=& \delta\omega_{2+} E_+ + \delta\omega_{2-} E_-
              = {1\over 2} \bar\partial J~ dz_-^2  \ .
\end{eqnarray*} 
This is the quadratic behavior of the lens equation near a critical point. 
The role of the critical direction is clear. 
If $z_\circ$ is a critical point, two images separated along the critical direction 
at $ z = z_\circ \pm |dz_-|E_-$ are from a same source. If $\omega_\circ$ is
the caustic point corresponding to $z_\circ$ under the lens equation, then the 
position of the source, $\omega = \omega_\circ + \delta\omega$, 
that produces the two images is 
such that the shift $\delta\omega$ is in the same direction as the gradient
of the Jacobian determinant $\nabla J = (J_+, J_-)$ at $z_\circ$. 
See Fig. 1.  We refer to this caustic point $\omega_\circ$ as the 
{\it preferred caustic point} for $\omega$.

\paragraph{\it \underline{Question :} \ \
Suppose $\omega$ is an arbitrary source position near the caustic curve. Then,
it is conceptually straightforward how to find the images of $\omega$ near the 
critical curve.}

\paragraph{\it \underline{Solution :} \ \
We find a caustic point $\omega_\circ$ such that  
$\delta\omega = \omega - \omega_\circ$ is parallel to $\bar\partial J$ 
(or $\nabla J$) at $z_\circ$ where $z_\circ$ is the critical point that 
corresponds to $\omega_\circ$.} Then, two images of $\omega$ can be found in 
the critical direction from the critical point 
$z_\circ$: ~$ z = z_\circ \pm |dz_-|E_-$ where
\opeqn
 |dz_-| = \sqrt{4\delta\omega_{2+}\over J_+}
        = \sqrt{4\delta\omega_{2-}\over J_-}
        = \sqrt{2\delta\omega\over \bar\partial J}  \ .
\label{eqVariety}
\cleqn
The two images have opposite parities and same
magnification as reflected in the signs and absolute value of the Jacobian
determinant at the image positions (rb99.11).
\opeqn
  J (z_\circ + dz) = dJ = dz_- J_- = \pm |dz_-| |J_-|
    = \pm\sqrt{4\delta\omega_{2-} J_-}
\label{eqJval}
\cleqn
If $\delta\omega$ is antiparallel to $\bar\partial J$, the inside of
the square roots in equation (\ref{eqVariety}) is negative, and $\omega$
generates no images near the critical curve.
\begin{enumerate}
\item
Once the directionalities of the 
image line ($\pm E_-$) and source line ($\bar\partial J$) are determined, 
the lens equation becomes a real symmetric quadratic equation from the image 
line to the source line. The lens equation from the image line to the source 
line has two solutions for ~$\delta\omega\partial J >0$ (real and positive), 
no solutions for ~$\delta\omega\partial J <0$ (real and negative),
and degenerate at the critical point where ~$\delta\omega\partial J =0$.  
It is exactly like the simple real quadratic equation $y= a x^2$:   
$y= a x^2$ has two solutioins for ~$a y> 0$, and no 
solutions for ~$a y<0$. At the critical point $x=0$ where $dy/dx=0$, 
the solutions are degenerate. 

{\it
It should be worth emphasizing that the image line ($\pm E_-$) and source line
($\bar\partial J$) are not orthogonal to each other
even though it may be tempting to imagine 
so because of our perception from the graph $y= a x^2$ which we are accustomed 
to draw in orthogonal grids.} 

\item
In fact, the second order approximation fails where the image line and the 
source line are orthogonal ($J_- =0$). Since 
equation (\ref{eqMinusOne}) fails where $J_-=0$, we need to consider the third 
order terms. The underlying reason can be seen in that the second order terms
accomodate only a migration along the critical curve as is clear from equation 
(\ref{eqJval}).

\item
Fig. 1 clearly shows that the larger the angle between $\nabla J$ and $E_-$, 
the more tangentially the images $z_\pm$ are shifted from $z_\circ$, and   
the smaller the normal distance of the images from the critical curve which 
determines the $J$-value ($dJ$). If we consider a small stellar disk near the
caustic curve, the images will be extensively elongated along the critical 
curve where $J_+ > J_-$.  

\item
Geometrically, $J_-=0$ means that the critical direction is tangent to the 
critical curve.  Let's consider the lens equation as a mapping restricted to 
the critical curve. Then, the restricted mapping is stationary or critical
where $J_-=0$. In other words, the caustic curve as an integral curve of the 
tangent develops a point (cusp) where $J_-=0$ because the ``speed" vanishes 
as in the trajectory of a pendulum\footnote{ A simple pendulum traces a radial
arc and may not convincingly mimick a caustic curve which is a closed curve with
interior (which can be heirachical). If we consider a Foucault pendulum, the
rotation of the earth should make it a closer analogy to the caustic
curve as a trajectory near a cusp if not too close to the equator. 
Nontheless, the directional change of the velocity at a stationary point is 
along the radial arc motion, which is the motion of a simple pendulum.} 
at the maximum gravitational potential. 
\opeqn
\int d\omega = \int d\omega_+ E_+ + d\omega_- E_- =
  \int_{critical\ curve} 2 dz_+ E_+
\cleqn
\opeqn
0 = dJ = dz_+ J_+ + dz_- J_- = dz_+ J_+  \quad\Rightarrow\quad  
\delta\omega_+ = 2 dz_+ = 0 
\cleqn
The cusps are the only singularities on the caustic curve. That is because the 
lens equation is smooth on the critical curve, and the critical curve is smooth. 
(A critical curve can have bifurcation points that are four-prong vertices, and 
the caustic curve develops cusps; see RBCPLX). 
Thus, the caustic curve is smooth except at the critical points where the 
restricted mapping is stationary.  

\item
If $J_+=0$, the two normals to the critical curve and the caustic curve
coincide. $|J_-| = |\nabla J|$ and $|\delta\omega_{2-}| = |\delta\omega|$.
The image line (the line connecting the two images) is perpendicular to the
critical curve, and $|\delta\omega|$ is the (shortest)
distance from $\omega$ to the caustic point onto which the critical point
that disects the image line is mapped under the lens equation. There is no
need to be an alarmist for that the second term in equation (\ref{eqVariety}) is
not well defined where $J_+=0$. Both $\delta\omega_{2+}$ and $J_+$ vanish, and
the second term is not sufficient to offer a finite number when $J_+=0$, but there
is no conceptual outrage. The last two expressions are
equivalent and offer the value of $|dz_-|$.   
\end{enumerate}

\paragraph{\underline
{\it What is $\delta$ in $J^{-1} \propto \sqrt{\delta^{-1}}$?} }

In this second order approximation, $J(\omega)$ is derived from two relations.
\opeqn
 J = J_- ~dz_- \ ; \qquad
    dz_- =  \pm\sqrt{2\delta\omega\over \bar\partial J}
\cleqn
The first equation holds since we chose the critical point $z_\circ$ such that
$z_\circ$ intersects the image line (along the critical direction). In fact,
$z_\circ$ bisects the image line. The second equation depends on (the square
root of) the ratio of $\delta\omega$ to $\bar\partial J$, and the ratio can be
expressed variously as shown in equation (\ref{eqVariety}). Thus, the title 
question ``What is $\delta$ in $J^{-1} \propto \sqrt{\delta^{-1}}$?" is not a
question well posed because the answer depends on the choice of the multiplication 
function. However, given the preferential appearance of $J_-$ in $J = dJ = J_- ~dz_-$
(since $dz_+ = 0$ here), we can agree to implicitly refer to $4 J_-$ 
as the multiplication function. Then, the ``truly relevant distance"
is $\delta = |\delta\omega_- (\omega_\circ)|$ and is shown in Fig. 1 
in comparison to the (shortest) distance to the caustic and 
``vertical distance"\footnote
{Given a source position $\omega$ and the caustic curve, the distance
of the source position to the caustic curve is intrinsically well defined.
However, a ``vertical distance" can be defined only when a ``vertical direction" 
is defined. In Fig. 6.1 of SEF and subsequently in Gaudi and Petters (2001), the
``vertical direction" is defined by the critical direction ($\pm E_-$) at a 
caustic point where the power series expansion is made. Since the choice of a 
caustic point as the origin of the power series expansion of the lens equation 
can be arbitrary, the ``vertical direction" is not uniquely determined for a
given source position $\omega$. That is in contrast with the fact that the
images and their magnifications are uniquely determined given a source position.
In other words, a ``vertical distance" of a source position is not an inherently 
meaningful quantity. In Fig. 1, the ``vertical distance" is depicted as the
distance from the source position $\omega$ to the caustic curve in the  
critical direction defined at $\omega_\circ$.} (to the cautic curve).

\paragraph{\it {\underline Answer:} \ \ 
If we choose $4 J_-$ as the multiplification factor, 
the distance $\delta$ is given by $|\delta\omega_{2-}|$ 
where $\delta\omega_{2-}$ is the normal component 
of $\delta\omega = \omega - \omega_\circ$ and the caustic point
$\omega_\circ$ is chosen such that $\delta\omega \parallel \bar\partial J$.}

\subsubsection*{\bf -1.1 \ \ 
Non-preferred Caustic Point as the Origin: $\omega_c \not= \omega_\circ$}  

If we choose an arbitrary caustic point $\omega_c$ in the caustic region of 
interest as the origin of a power expansion, then
$\delta\omega = \omega - \omega_c$ where $\omega$ is a source position
in the caustic region. 
If we let $x \equiv \delta\omega_+$ and ~$y \equiv \delta\omega_-$; 
~$u \equiv \delta z_+$ and  ~$v \equiv \delta z_-$; 
~$a \equiv J_+$ and ~$b \equiv J_-$ for visual simplicity, then, 
the second order lens equations (\ref{eqLomega}) in the neighborhood 
of $\omega_c$ are rewritten as follows.
\begin{eqnarray}
 x - 2u &=& - {1\over 4} \left( a (u^2-v^2) + b ~2 u v \right) \\
 y      &=& - {1\over 4} \left( b (u^2-v^2) - a ~2 u v \right)   
\label{eqxy}
\end{eqnarray}
One combination of the two equations reads as follows.
\opeqn
 (bx - ay - 2bu) = - {1\over 4} (a^2 + b^2) ~2 u v
\cleqn
Since we have chosen a non-preferred caustic point, the image position solutions
will include shifts along the critical curve from the origin $z_c$  
where $\bar\partial J$ is calculated.  
In this second order approximation, the change of $J_\pm$ along the critical
curve can not be incorporated because higher order derivatives of the Jacobian
determinants become involved in the third order terms and higher. Thus, 
$\nabla J$ is implicitly considered constant in the neighborhood of $z_c$.  

\paragraph{\underline
{\it Linear Approximation for $dz_+$:}} \ \

We take the linear approximation as in rb99.12 (but correctly) by ignoring the RHS.
\opeqn
 (bx - ay) = 2 b u
\label{eqLinear}
\cleqn
In rb99.12, the second term $(-ay)$ in equation (\ref{eqLinear}) is missing. 
This missing term causes what amounts to an intrinsic
violation of the second order nature of the equations or of the proper handling
of a triangular matrix. Since the linear contribution for $y$ vanishes, the 
second order contribution to $y$ feeds back to $x$ through the second order 
contributions. In order to cut at the linear order of $\delta\omega_\pm$,
we need to fully incorporate these contributions in the second order in $dz_\pm$.
For example, if $x$, $y$, and $u$ are an order of $10^{-3}$ as in a typical case 
of a main sequence source star in the Galactic bulge lensing, then $v$ is an order
of $\sqrt{10^{-3}}$ and $\delta\omega_{2+}$ is of the same order as 
$\delta\omega_{1+}$. In fact, we know that we can choose $\omega_\circ$   
for the caustic point such that $\delta\omega_{1+}$ vanishes while 
$\delta\omega_{2+}$ doesn't. Once $u$ is determined by the linear equation
in (\ref{eqLinear}), we can find $v$ in terms of $u$ and $y$ from equation 
(\ref{eqLinear}), which is the same as rb99.13 with $dz_+$ reinstated. 
\opeqn
 v = -{a\over b} u \pm 
   \sqrt { {a^2+b^2\over b^2} u^2 + {4 y \over b} }  
\label{eqVval}
\cleqn 
The first term in the RHS describes a shift along the critical curve, and
so the non-vanishing contribution to $J = dJ = au+bv$ comes from the second 
term in the RHS. 
\opeqn
 J =  \pm b \sqrt {{a^2+b^2\over b^2} u^2 + {4 y \over b} }
\label{eqJdJ}
\cleqn 
We can rewrite it in terms of the original notations.
\begin{eqnarray}
 J (\omega) &=& \pm \sqrt{ 4 J_- \Delta\omega_- } \\
    \Delta\omega_- &\equiv& \delta\omega_- + {(\nabla J)^2\over 16 J_-} 
     \left( \delta\omega_+ - {J_+\over J_-} \delta\omega_- \right)^2
\label{eqJdJtwo}
\end{eqnarray}

If we stare at the solutions in (\ref{eqLinear}) through (\ref{eqJdJtwo}) for
a moment, a few things are clear.
\begin{enumerate}
 \item
  The line (or graph) defined by $(bx-ay) = 2 b u$ in equation (\ref{eqLinear})
  is parallel to  $\nabla J$. Thus, the linear approximation for $dz_+$ in 
  equation (\ref{eqLinear}) implies that the line $(bx-ay) = 2 b u$ should 
  connect the source position $\omega$ and the approximate preferred caustic
  point. The approximate preferred caustic point is marked as $w-$ in 
  Fig. \ref{fig-approx}. The line $(bx-ay) = 2 b u$ intersects the $x$-axis 
  at $x= 2u$ and is at a distance $d$ from the line defined by $(bx-ay) = 0$ 
  as shown in the same figure. 
\opeqn
  d = {2 u \over \sqrt{a^2 + b^2}} =  {| 2 dz_+| \over |\nabla J|}
    = {1\over |\nabla J|}
       \left| \delta\omega_+ - {J_+ \over J_-} \delta\omega_- \right| 
\cleqn
 \item
The approximate preferred caustic point determined by the line  
$(bx-ay) = 2 b u$ can be found by solving equations (\ref{eqVval}) and 
(\ref{eqQcaustic}) simultaneously.
\begin{eqnarray*} 
 x &\approx& 2u  - {a (a^2+b^2)\over 16 b^2}~(2u)^2 ~\approx~ 2u   \\ 
 y = - {a^2+b^2\over -16b} x^2 &\approx& {a^2+b^2\over 4b} u^2
    \approx {a^2+b^2\over 16 b}
    \left( \delta\omega_- - {J_+\over J_-}\delta\omega_+^2 \right)^2
\end{eqnarray*}
The approximations for $x$ and accordingly for $y$ have been made based on the 
validity condition for $x$ in equation (\ref{eqXrange}). The caustic point is
marked by $w0$ in Fig. (\ref{fig-approx}) and will be referred to as
$\omega_{c_0}$ in this text.   \\

\paragraph{\underline {\it What is $\Delta\omega_-$?}} \ \

If we consider $\delta\omega(\omega_{c_0}) \equiv \omega - \omega_{c_0}$, then 
$\Delta\omega_- $ in equation (\ref{eqJdJtwo}) is nothing but the {\it normal 
component} of $\delta\omega(\omega_{c_0})$.   
\opeqn
  (\omega - \omega_{c_0})\cdot E_- = (\delta\omega_- - y)  
               ~\approx  ~\delta\omega_- 
    + {a^2+b^2\over 16 b}
    \left( \delta\omega_- - {J_+\over J_-}\delta\omega_+^2 \right)^2
\cleqn
We impose an implicit understanding that the caustic point 
$\omega_{c_0} = \omega (J=0)$ is to be found in the direction of the normal  
$\bar\partial J$ calculated at $z_c$ where the power series expansion    
of the lens equation is made. 
\opeqn
  \Delta\omega_- = (\omega - \omega(J=0))\cdot E_-
          = \delta\omega_- - \delta\omega_-(J=0)
\cleqn
If $\omega_c$ coincides with $\omega_{c_0}$,
then $\omega_c = \omega_{c_0} = \omega_\circ$, and
$\Delta\omega_- = \delta\omega_-$.

\item
  The causic curve in the second order approximation satisfies equations 
  in (\ref{eqCpoint}), and the resulting quadratic equation for the caustic 
  curve is written in equation (\ref{eqCaustic}). If $(x, y)$ is a caustic 
  point, then $x$ and $y$ lie on the line defined by the first equation in  
  (\ref{eqCpoint}), and the line intersects the $x$-axis at $x = 2u$ as       
  shown in Fig. \ref{fig-approx}. The caustic point for the given value of
  $u$ differs from the caustic point determined by the line $(bx-ay) = 2bu$,
  and the former is marked as $w+$ in Fig. \ref{fig-approx}. It should be
  clear that $w0$ is always located between $w+$ and $w-$. The linear 
approximation for $dz_+$ in equation (\ref{eqLinear}) is valid when 
$w+ \approx w0 \approx w-$ with the second order approximation. 

\item
\paragraph{\underline {\it Error from Nonlinear Correction for $dz_+$:}} \ \
In order to test the goodness of the linear approximation for $dz_+$, let's
examine the intersection point of the second order equation (\ref{eqxy})
with the $x$-axis. If we set $y=0$, equations of (\ref{eqxy})
leads to the following.
\begin{eqnarray*}
 v &=& {1\over b} (-a \pm \nabla J) u \ ;  \\
 x &=& 2 u + {a^2 + b^2 \over 8 b^2} (a \pm |\nabla J|)~ (2 u)^2 
\end{eqnarray*}
As we mentioned before, $u$ is sufficiently smaller than 1 (Einstein ring
radius), and the second term can be ignored unless it is near a cusp
where $b = J_- \approx 0$ or $|\nabla J|$ is large as in the case of small
caustics such as planetary caustics we will briefly discuss below. 
\end{enumerate}

\paragraph{\underline {\it Quadratic Caustic Equation:}} \ \
Now, we write out the quadratic caustic equation and its validity used
in the previsous section.
The equation for the caustic curve can be obtained from equation (\ref{eqxy})
and the condition $0 = dJ = J_+ dz_+ + J_- dz_- = a u + b v$.    
\opeqn
 -16 b y = {a^2 + b^2\over b^2} (bx + ay)^2 \quad \Rightarrow \quad
  bx = - ay \pm \sqrt{-16 b^3 y \over a^2 + b^2}  
\label{eqCaustic}
\cleqn
\begin{enumerate}
 \item The equation is non-linear in $y$, but we can confirm that the origin 
       ($x=y=0$) is indeed a critical point ($dy/dx=0$). 
 \item Where $by < 0 $, every $y$ has two solutions. See Fig. 4.  
 \item The slope $dy/dx$ diverges where the ``source line" (parallel to $\nabla J$)  
intersects with the caustic curve given by equation (\ref{eqxy}).
\opeqn 
  y_\infty = - {4 b^3\over a^2 (a^2 + b^2) } \quad \Rightarrow \qquad
  b x_\infty - a y_\infty = 0 \ .  
\cleqn  
This indicates where equation (\ref{eqxy}) fails to describe the 
original caustic curve. We can say that the quadratic equation (\ref{eqxy})
(not necessarily quadratic in the orthogonal coordinate variables $x$ and $y$)
is valid where  $|y| << |y_\infty|$, 
or equivalently where $(bx)$ dominates over $(-ay)$.  
From this condition, we get a criteria for $|x|$.  
\opeqn
  |x| <<  {4 \over |J_+|}  \left( J_- \over \nabla J \right)^2 
\label{eqXrange}
\cleqn
For small $x$ where $|ay| << |bx|$, equation (\ref{eqxy}) becomes a simple
quadratic equation.
\opeqn
 -16by = (a^2 + b^2) x^2 \quad \Leftrightarrow \quad 
 - 16 J_- \delta\omega_- = |\nabla J|^2 \delta\omega_+^2
\label{eqQcaustic}
\cleqn     
\item  
If $J_-=0$, the range $|x| << 0$. In other words, the quadratic equation
         for the caustic curve is completely invalid at the cusp. It is 
         consistent with that the second order approximation of the lens 
         equation fails where $J_- \approx 0$.   

If $J_- =0$, the source line $bx-ay=0$ becomes $y=0$, and so 
         $dx/dy =0$ at $y=0$. This is another way to look at the failure of
         the quadratic equation near a cusp. 
\item
If $J_+ =0$, the range of $|x|$ in equation (\ref{eqXrange}) is unlimited.
That is directly because $(-ay)$ term vanishes and so $bx$ term dominates. 
If we look at Fig. 4, $bx+ay=0$ and $bx-ay=0$ become degenerate, and the 
quadratic caustic curve in Fig. 4 becomes symmetric: $bx = \pm f(y)$,
or $-16by = (\nabla J)^2 x^2$. 
\end{enumerate}

\section { A Single Lens with Constant Shear } 

In 1979, the first double quasar Q0957+561A,B (Walsh, Carswell, and Weymann 1979) 
was discovered and interpreted as two images of one quasar gravitationally lensed
by a galaxy (or a group of galaxies) as a continuum distribution of mass. 
Chang and Refsdal (1979) considered the effect of ``image splitting" of an image
due to the granularity of a single star which happens to be {\it near} the 
astrometric (or transverse) position of the image.  
The angular scale of the ``image splitting" 
of an image of a quasar (at a cosomological distance) by a stellar mass object 
is an order of microarcsecond while the two images Q0957+561A,B are separated by
$\sim 6$ arcseconds. Thus, the {\it nearity} of the would-be image lensed by the 
galaxy to the (microarcsecond) lensing star amounts to the precision level of   
$\sim 10^{-6}$, and the gravitational influence of the lensing galaxy in the 
small area defined by the microarcsecond lensing radius of the star can be 
considered constant. 

The equation for a single point lens with constant shear entails a preferred 
direction and that is the direction of the (effective) mass which generates the
constant shear in the small area of interest around the single point mass.   
In other words, {\it constant shear} is intrinsically an approximation 
and can be obtained by taking large mass and large distance limits of more
physically well-defined mass distributions.

As a concrete example, let's take a large mass parity limit of the binary lens 
in Eq. (\ref{eqLens}). The lens equation has been normalized by $R_E$ 
the Einstein ring radius of the total mass, and we can suppose that
$\epsilon_1 << \epsilon_2$ such that $\epsilon_1$ is the mass fraction of the 
(microarcsecond) lensing star.  If we rescale the lens equation 
by the Einstein ring radius of the stellar mass $\epsilon_1$, then $\epsilon_1=1$, 
and $\epsilon_2 >> 1$.  We can set $x_1=0$ (choice of the coordinate origin), 
and then the microarcsecond lensed images will be around $|z| \sim 1$.  In order
for the lensing by the larger mass not to dominate the lensing behavior of the 
images around $|z| \sim 1$, the bending angle by the larger mass must be the 
order of the bending by the stellar mass $\epsilon_1$ or smaller. For images 
around $|z| \sim 1$, $|x_2| >> |z|$, and the lens equation can be rewritten
as follows.     
\opeqn
\omega - {\epsilon_2\over x_2} = z - {\epsilon_1\over \bar z - x_1}
       + {\epsilon_2 \bar z \over x_2^2} + {\cal O}(x_2^{-3}) \ . 
\label{eqShear}
\cleqn 
We note that $\sqrt{\epsilon_2} \sim R_E$ is the Einstein ring radius of the 
mass $\epsilon_2$ as a single lens. If $\epsilon_1$ is at a distance of 
$\sim R_E \approx \sqrt{\epsilon_2}$ from $\epsilon_2$, then the source at
$\omega \approx \epsilon_2/ x_2$ produces images at whose positions the 
influence of the large mass $\epsilon_2$ can be considered a constant shear
of order 1: \ $\gamma \equiv \epsilon_2 / x_2^2 \sim {\cal O}(1)$. 

The constant shift $\epsilon_2/ x_2$ indicates that the emission source is
near the position of the large mass $x_2 >> 1$ for the images around with 
$|z| \sim {\cal O}(1)$, and the effect of the single point mass is to ``split" 
the would-be image around the Einstein ring of the large mass. The {\it outer
image} is splitted by a quadroid and the {\it inner image} is splitted
by two trioids. These behavior can be very easily understood from the 
binary lens equation which is a physically well-defined closed system.

It is customary to ignore the {\it constant shift} $\epsilon_2/ x_2$ in a 
lens equation with a {\it constant shear}.
\opeqn
\omega = z - {1\over \bar z} + \gamma \bar z  
\label{eqLeqShear}
\cleqn
This equation is often referred to as Chang-Refsdal equation (e.g., SEF) and 
the constant shear term points to the positive direction of the $x$-axis.
This particular directionality has been inherited from 1) the binary
equation (\ref{eqLens}) where the lens axis is along the real axis 
and 2) the implicit assumption in the derviation of 
equation (\ref{eqShear}) that the larger mass was in the positive direction.  
In general, the constant shear coefficient $\gamma$ would be a complex number 
$\gamma \equiv \epsilon_2 / \bar x_2^2$ half whose phase angle points to the 
effective direction of the larger mass that generates the constant shear. 

It is noteworthy that the second term in equation (\ref{eqLeqShear}) due to 
the (microarcsecond) lensing star is also a shear term, and this shear term 
vanishes at $\infty$ -- far away from the mass that generates the shear. However, 
the constant shear does not vanish at $\infty$. It is in a sense a reminder that 
constant shear is inherently an expression of approximation, and the lens
equation with a constant shear should be treated as an incomplete equation
which requires auxiliary assumptions or interpretations to be a physically
viable model for a lensing system. For example, the ``source
position" in equation (\ref{eqLeqShear}) can not be the position of the source
because the ``source position" is in practice the position of the would-be 
image generated by the larger mass that effects the constant shear around the 
(microarcsecond) lensing star. 

In order to see the effect of the constant shear, let's remove the point mass 
of the star from equation (\ref{eqLeqShear}). 
\opeqn
 \omega = z + \gamma \bar z \ ; \qquad J = 1 - |\gamma|^2
\cleqn 
The equation is linear, and the constant shear generates only one image.
The magnification ($|J|^{-1}$) and parity (sign of $J$) of the image is the 
same everywhere irrelevantly of the position of the ``source", which is 
arguably the essence of a {\it constant shear}.
\opeqn
 z = {\omega - \gamma\bar\omega \over 1-\gamma^2}
\cleqn   
The image is contracted in the direction of the shear $\gamma$ by a factor
$(1+\gamma)^{-1}$ and elongated in the orthogonal direction by a factor
$(1-\gamma)^{-1}$.  The distortion is universal where the constant shear 
approximation applies. In weak lensing where the shape of the objects  
is one of the measurable and interpretable quantities, 
such systematic distortions due to lensing can be detected and are being 
pursued to study large scale structures that may shed light on the history and
make-up of the universe including dark stuff
(matter, energy, essence, extra dimensions, topological defects, ... , 
even though it is unclear whether the essence is essential, extra dimensions 
are most likely a fundamental ingredient but for the time being a tantalizing 
yet unproven conjecture, and the exact nature of the topological defects 
or objects may turn out to be as fascinating as the existential properties of 
the space or the baryon number and remain illusive for a long time to come). 
How well one can determine the shear distribution crucially depends on the 
shape resolution and statistics of the shape objects. 
A degree scale large format space imager Galactic Exoplanet Survey Telescope
will be a fitting explorer even though the wings have been clipped again.

\subsection { The $\kappa$ Field of the Single Lens with Constant Shear }

The linear differential behavior is insensitive to the constant
shift of the source position and can be discussed routinely as usual.
As in the case of $n$-point lenses \citep{rh97,limb,quadlens}, the linear 
differential behavior of the lens equation (\ref{eqShear}) is completely 
determined by one analytic function $\kappa$, and so we use and follow the facts, 
conventions, and analysis patterns in the references (RBCPLX from here on). 
The Jacobian of the lens equation 
can be written as follows where the $2\times 2$ matrix indicates that the
underlying space of the complex plane is a 2-d linear space.    
\opeqn
   {\cal J} = \pmatrix{1 \  \bar\kappa \cr
                       \kappa \  1 } \ ; \qquad
           \kappa \equiv {\partial\bar\omega}
               \equiv {\partial\bar\omega\over\partial z} 
              = {1\over z^2} + \gamma
\label{eqJacobian}
\cleqn
The condition for criticality, $|\kappa|=1$, is a second order analytic
polynomial equation, and so the phase angle of $\kappa$ changes by $2\pi$
along the critical curve. 
\opeqn
   \kappa = e^{i2\varphi} \ ; \qquad \Delta\varphi = 2\pi
\label{eqCritKappa}
\cleqn
Thus, the total topological charge of the critical loops or caustic loops 
is $\sum |e| = 1$, and the caustic is made of one 4-cusped quadroid or
two 3-cusped trioids. The depiction can be found in \citet{ch84} and also
in SEF.Fig.8.8. There are two limit points (where $\kappa = 0$) on the 
imaginary axis (orthogonal to the direction of the constant shear) at 
$z = \pm i /\sqrt{\gamma}$, and the trioids enclose the limit points.    
\opeqn
   J(z) = 1 - |\kappa (z)|^2  \le  1 \  ; \qquad J(\infty) =  1 - \gamma^2
\cleqn
The Jacobian determinant $J(z)$ of this lens equation with a constant shear
is bounded by $J(z) \le 1$. The maximum holds at the finite limit points 
$z = \pm i/\sqrt{\gamma}$ but not at infinity. In fact, the critical curve can 
pass through the infinity and the image at infinity can be infinitely bright. 
This unphysical behavior derives from that the lens equation with a constant
shear is accommodated by an infinitely large mass at infinity when the lens 
equation is extended to the whole lens plane.  
This behavior is easy to understand when seen as an approximation of the
binary lens. The critical points at infinity correspond to the bifurcation
points where the topology of the critical curve changes from one loop 
to two loops. At the same time, the topology of the caustic curve changes 
from one quadroid to two trioids.

\subsection{ Images }  

We have derived the lens equation (\ref{eqLeqShear}) of a single lens with a
constant shear from the binary lens
equation and can read off some of the basic properties of the equation 
following the recipes developed for binary (or $n$-point) lenses briefly
reviewed in the following section.
\begin{enumerate}
\item
 The caustic loops are simple loops. In other words, they neither self-intersect
nor nest.  Thus, the source plane accomodates only two domains: 
${\cal D}^\circ$ (outside) and ${\cal D}^1$ (inside).    
\item 
 There are two images outside the caustic loop(s). One only
needs to examine the number of images of $\omega = \infty$  because the
outside domain ${\cal D}^\circ$ includes $\infty$.  
\item
 A source inside a caustic loop (${\cal D}^1$) produces four images, and the 
caustic loop itself produces three image loops one of whose is the critical 
curve. The caustic curves have cusps, and so the two ``non-critical curves" have
kinks at the corresponding points unless they are precusps.  However, the 
critical curve is smooth because the inverse mapping of the lens equation from 
the caustic curve to the critical curve is singular at the cusps
the singular points. Precusps are bifurcation points of the critical curve
and ``non-critical curves" (see Fig.10 in \citet{quadlens}).     
\end{enumerate}      

The constant shear breaks the axial symmetry of the single lens but leaves
a residual reflection symmetry as is familiar from binary lenses \citep{binary}. 
If the source is on the lens axis, $\omega = \bar\omega$, there are two images
on the the lens axis, $z = \bar z$, and  two extra images on a circle
$|z|^{-1} = \sqrt{1-\gamma}: \gamma > 1$. The lens equation subject to  
$\omega = \bar\omega$ leads to the following equation.
\opeqn
 0 = (z-\bar z) \left(1-{1\over |z|^2} - \gamma \right)  
\cleqn
\begin{enumerate}
 \item
\underline{ {\it $z =\bar z$:}} \ \  
The two images on the lens axis are very much like the two images of a single 
lens that form on the line defined by the positions of the lens and source. 
\opeqn
 \omega = z(1+\gamma) - {1\over z} \quad \Longrightarrow \quad 
\left[{\omega\over\sqrt{1+\gamma}}\right] = \left[\sqrt{1+\gamma}~z \right] 
 - {1\over \left[\sqrt{1+\gamma}~z \right]}
\cleqn
The second equation shows the correspondence with the equation of a single lens.
\item
\underline{{\it $|z|^{-2} = 1-\gamma \ (>0)$:}} \ \ The two images
are on the circle of radius $|z| = 1/\sqrt{1-\gamma}$ centered at the lens 
position (here the origin) where the real component of the image positions
is given by a simple linear scaling of the source position from the lens.
\opeqn
 \omega = 2 \gamma ~Re(z) \ : \qquad \gamma > 1 \ \ {\rm and} \ \   
          |Re(z)| \le {1\over \sqrt{1-\gamma}} 
\label{eqSaxisShear}
\cleqn 

When $\gamma < 1$, the caustic curve is a quadroid intersecting the lens axis.
The cusps of the quadroid are easily calculated: \ two are on the real axis 
(the lens axis) and has $\kappa = 1$, and the other two are on the imaginary 
axis and has $\kappa = -1$.
\opeqn
 \omega_{cusp} = {\pm 2\gamma \over \sqrt{1-\gamma}} \ , \qquad
                     {\pm i~ 2\gamma \over \sqrt{1+\gamma}} 
\label{eqCuspShear}
\cleqn
The two images on the circle exist when $\omega$ is inside the quadroid, which 
can be easily read off from equations (\ref{eqSaxisShear}) and
(\ref{eqCuspShear}). 
The Jacobian determinant of the images on the circle is given as follows.
\opeqn
 J = 4 \gamma (1-\gamma) \sin^2 \theta \ \ge 0 \  : \qquad 
       z = {e^{i\theta}\over\sqrt{1-\gamma}}
\label{eqJringShear}
\cleqn
The images on the circle are positive images except at the
cusps. At the cusps on the real axis given in equation (\ref{eqCuspShear}),
$J=0$, and the circle is tangent to the critical curve. We can easily check that
the images on the real axis are inside the critical curve and so are negative
images. The total parity of the images of a single lens with a constant shear
is zero.

When $\gamma > 1$, the caustic curve is made of two 
triods enclosing the two limit points, and the two caustic loops are off
the lens axis. Thus, $\omega = \bar\omega$ has only two images and they 
have opposite parities exactly as in the case of a single lens.   
\end{enumerate}

The fact that the two extra images are on a circle deserves a modest attention.
If we decrease the effect of the constant shear, $\gamma \rightarrow 0$, the
circle approaches the critical curve of the single lens of radius 1, namely
the Einstein ring radius: \ $J \rightarrow 0$ and $|z| \rightarrow 1$.
In general, Einstein ring, critical curve, and image ring are all distinct.
Einstein ring determines the (transverse) distance scale of lensing. Critical
curve is where the Jacobian determinant of the lens equation vanishes, 
images are degenerate, and the magnitude of the images diverge. Image
ring is a proxy ring which a finite size source fallen inside a caustic loop
produces, and its shape and intensity depends on the caustic structure and the 
relative position of the source. In the case of a single lens, they all
converge to one ring. In the case of the single lens with a constant shear,
its proximity to a single lens makes the image ring quite circular. In fact,
the same line of thought offers a valid intuition for caustics of higher
multiple lenses when the effect of the lens elements but one can be considered
a long distance effect.

\section{ Binary Equation, Quasi-Analytic Equations }

The binary lens equation is written as follows (see RBCPLX, references 
therein, or others). 
\opeqn
  \omega = z - {\epsilon_1\over \bar z - x_1}
             - {\epsilon_2\over \bar z - x_2} \ ,
\label{eqLens}
\cleqn
where $\omega$, $z$, and $x_j: j =1,2$ are the positions of a source, an image,
and the lens elements of fractional masses $\epsilon_j: j =1,2$ in the two
dimensional sky as a complex plane. We have chosen the lens axis to be along
the real axis so that $x_1$ and $x_2$ are real, and the unit distance scale is
given by the Einstein ring radius $R_E$.
\opeqn
  1 = R_E = \sqrt{4GMD} \ ; \qquad
   {1\over D} = {1\over D_1} + {1\over D_2}
\cleqn
where $D$ is the reduced distance. 

When there are $n$ point lens elements where $n > 2$, we only need to extend
the range of the index to $j = 1, 2, ... , n$ and replace the lens position  
vectors to $\bar x_j$ because we can not line them up all on the real axis. 
The form of the Jacobian matrix in equation (\ref{eqJacobian}) applies to   
any quasi-analytic lens equation, and $\kappa \equiv \partial \bar\omega$
is easily calculated for each class of lens equations.  
\begin{eqnarray*}
 \kappa  & = & \sum_j {\epsilon_j\over (z - x_j)^2} \\
   J & = & 1 - |\kappa|^2 \le 1 \  ; \qquad  J(\infty) = 1    
\end{eqnarray*}

\subsection { Eigenvalues, Eigenvectors, and Orthogonal Decompositions }

The quasi-analytic lens equations 
-- the lens equation is not analytic, but the differentials are -- have 
been discussed in RBCPLX and is reviewed here briefly.
The eigenvalues of the Jacobian matrix are easy to find,
and $\lambda_-$ vanishes on the critical curve
($|\kappa|=1 \Leftrightarrow J =0$).
\opeqn
  \lambda_\pm = 1 \pm |\kappa|  \ ,
\cleqn
If $2\varphi$ is the phase angle of $\kappa$ such that
\opeqn
\kappa \equiv |\kappa| e^{2i\varphi} \ ,
\cleqn
the eigenvectors are given by $(\pm) e_\pm$ where we choose the 
basis vectors $e_\pm$ as follows. 
\opeqn
 e_+ \equiv {e^{-i\varphi} \choose e^{i\varphi}} \ , \quad
 e_- \equiv {i~e^{-i\varphi} \choose -i~e^{i\varphi}}  \ ;
 \qquad ||e_\pm|| = \sqrt{2} \ ,
\label{eqEvectors}
\cleqn

The orientation of the basis vectors defined here can jump or mismatch   
around the critical curve. Here we are mostly concerned with the local
behavior of the images near the critical curve, and the global 
orientability of the basis vectors as chosen in (\ref{eqEvectors}) along
the closed curves would not be a relevant concern. 

The complex plane is a real plane with complex structure, 
and it is sufficient to write out half the 2-d equation 
because the other half is the complex conjugate of the former half.    
\opeqn
  d\omega = dz + \bar\kappa d\bar z 
      + {1\over 2}\bar\partial \bar\kappa d\bar z^2 + {\cal O}(|dz|^3)
\label{eqDomega}
\cleqn
We only need to consider the upper components of the basis vectors,
which are mutually orthonormal complex numbers.  
\opeqn
 E_+ \equiv e^{-i\varphi} \quad ; \qquad 
 E_- \equiv i e^{-i\varphi}  
\cleqn   
$(E_+, E_-)$ forms a right-handed coordinate frame, and any vector can 
be decomposed into the orthogonal components.  For example, 
$dz = dz_+ E_+ + dz_- E_-$ where the linear coefficients $dz_+$ and 
$dz_-$ are given by the inner product $(dz, \bar E_\pm)$ 
that is uniquely defined from the norm of a complex number. 
If $dz = |dz| e^{i\theta}$ and $r = |dz|$,  
\begin{eqnarray} 
  dz_+ & = & Re (dz \bar E_+) = r \cos(\theta + \varphi) \equiv rC \\ 
  dz_- & = & Re (dz \bar E_-) = r \sin(\theta + \varphi) \equiv rS 
\label{eqDecomp}
\end{eqnarray}
We note that the decomposition coefficients are real.

We recall from RBCPLX that $E_+$ is always tangent to the caustic curve and 
so $E_-$ is always orthogonal to the cautic curve. Thus, decomposing the 
quantities relevant in lensing into orthogonal components in $E_+$ and
$E_-$ is a bit more than an idle exercise. The Jacobian determinant 
of the lens equation $J(z) = 1 - |\kappa|^2$ is a real function and
its differential $dJ$ can be calculated in any basis we are pleased to
choose. If the 2-d real plane is parameterized by $(\xi_1, \xi_2)$ such that 
$z = \xi_1 + i \xi_2$ and $\bar z = \xi_1 - i \xi_2$, then 
$\partial \equiv \partial_z = (\partial_1 - i \partial_2 )/2 $
and $\bar\partial \equiv \partial_{\bar z} = (\partial_1 + i \partial_2 )/2$.   
We may call these coordinate changes complexification and express 
$z = (\xi_1, \xi_2)^{\bf C}$ and $2 \bar\partial = (\partial_1, \partial_2)^{\bf C}$. 
\begin{eqnarray*}
 dJ & = & \partial_1 J d\xi_1 + \partial_2 d\xi_2     \\  
    & = & \partial J dz + \bar\partial J d\bar z   \\
    & = & \partial_+ J  dz_+ + \partial_- J dz_-  
\label{eqDJ}
\end{eqnarray*}
The gradient $\nabla J = (\partial_1 J, \partial_2 J) $ points in the 
direction of the maximum increase of $J$ and is a normal vector to the
critical curve. So is $\bar\partial J$.
\opeqn
  (\nabla J)^{\bf C} =  2 \bar\partial J
\cleqn 
From equation (\ref{eqDJ}) and conversion between $(dz_+, dz_-)$
and $(dz, d\bar z)$, we can confirm that $\partial_+ J$ and $\partial_- J$
are $E_\pm$ components of the normal vector $(\nabla J)^{\bf C}$.  
\begin{eqnarray}
 2\bar\partial J & = & \partial_+ J E_+ + \partial_- J E_-  \\
                 & \equiv &  J_+ E_+ + J_- E_- 
\label{eqJdcomp}
\end{eqnarray}
We remind that $J_+$ and $J_-$ are real, and so are $\partial_+$ and 
$\partial_-$. 
\opeqn
 \partial_\pm = E_\pm \partial + \bar E_\pm \bar\partial 
\label{eqDplusminus}
\cleqn
From here on (and retroactivley if applicable) we freely interchange between
$\nabla J$ and $2\bar\partial J$ because they are one and the same vector
expressed in different coordinates.

\subsection { Near the Critical Curve and Caustic Curve}

The variation equation (\ref{eqDomega}) can be sorted out into $E_\pm$
components.  In the linear order in  $r=|dz|$,
\opeqn
 d\omega = rC (1 +|\kappa|) E_+ + rS(1 - |\kappa|) E_-  \ ,
\cleqn
and the $E_-$-component vanishes on the critical curve because
$|\kappa| =1$. This vanishing linear component is behind the quadratic
behavior of the lens equation in the critical direction ($E_-$) in the
neighborhood of the cirtical curve and is at the foundation of the
{\it square-root-of-the -distance} dependence of the magnification of
the images near the critical curve. We have discussed fully in section -1 that 
the lens equation becomes a real symmetric quadratic equation from the image line
($\pm E_-$) to the source line ($\bar\partial J$) where $J_ \not\approx 0$.   

We calculate the second order term 
in equation (\ref{eqDomega}) at the critical point.   
\opeqn
  {1\over 2}\bar\partial \bar\kappa~ d\bar z^2
   = -{1\over 4} r^2 \left( (J_+ C_2 + J_- S_2) E_+ 
                    ~ + ~ (J_- C_2 - J_+ S_2) E_- \right)  \ , 
\cleqn
where $C_2 \equiv\cos 2(\theta+\varphi)$ and $S_2 \equiv\sin 2(\theta+\varphi)$.  
The $E_\pm$-components of the linear deviations and second order deviations 
can be listed as follows.
\begin{eqnarray}
 \delta\omega_{1+} & = & 2 rC  \\
 \delta\omega_{1-} & = & 0  \\
 \delta\omega_{2+} & = & - {1\over 4} r^2 (J_+ C_2 + J_- S_2) \\
 \delta\omega_{2-} & = & - {1\over 4} r^2 (J_- C_2 - J_+ S_2)  
\label{eqLomega}
\end{eqnarray}
If we consider a unit vector ${\bf unit} \equiv (C_2, ~S_2)$, then   
$\delta\omega_{2+}$ depends on the ${\bf unit}$-component of $\nabla J$
(inner product, or dot product), and $\delta\omega_{2-}$ depends on the signed 
area (rotation, or cross product) defined by ${\bf unit}$ and $\nabla J$. 
It is usually the case\footnote
{The gradient vanishes on the critical curve only where the critical curve
bifurcates and changes its topology. In the case of a single lens with 
constant shear $\gamma$, there are two bifurcation points at $\infty$ 
when $|\gamma|=1$. If we go back to the derivation from the binary lens 
equation, $|\gamma|=1$ holds when the constant shear is calculated on the 
Einstein ring of the larger mass.} that $\nabla J \neq 0 $.  The third order
terms have the similar structure as the second order terms. 
\begin{eqnarray}
 \delta\omega_{3+} & = & - {1\over 24} r^3 \left( (J_{++}-J_{--}) C_3 
            + 2 J_{+-} S_3 \right) \\
 \delta\omega_{3-} & = & - {1\over 24} r^3 \left( 2 J_{+-} C_3 
            - (J_{++}-J_{--}) S_3 \right)  
\label{eqThird}
\end{eqnarray}
The reason is because $\kappa$-field is analytic and so is orientable.  
In the third order, the phase is multiplied three times, and so the unit vector
is replaced by ${\bf unit} \equiv (C_3, ~S_3)$.

\subsection { Rough Estimation of $|\nabla J|$ }

In a typical Galactic bulge lensing where a star in (the direciton 
of) the bulge is lensed by a (faint) star or a planet system in the bulge or 
the disk, the stellar radius in units of the Einstein ring radius of the 
total mass of the lensing system is $r_\ast \sim 10^{-3} - 10^{-2}$.           

In order to get an idea of the magnitude of the gradient $J$ on critical 
curves, let's consider the simplest case of a single lens. For a single lens,
$\kappa = 1/z^2$; \ $J = 1- 1/|z|^4$; \ $|\nabla J | = 4/|z|^5$. On the 
critical curve, $|z|=1$, and $|\nabla J | = 4 >> r_\ast$. The gradient $J$ is
normal to the ring $|z|=1$ and so is the positive eigenvector $(\pm) E_+$. 
Thus, $J_-=0$, and $J_+ = (\pm) 4$. The fact that $J_-=0$ everywhere on the
critical curve underlies that the point caustic is a degenerate cusp.   

In a binary lens where its lens elements masses are comparable, 
we expect a similar range of values of $|\nabla J|$.  
We have shown the numerical values of $\sqrt{|J_-|}$ on the 4-cusped
central caustic of a binary lens with the mass ratio of $\sim 2 : 1$ in 
Fig. 1 of \citet{limb} (rb99). We also stated that trioids have somewhat
larger values. The larger the value of $|\nabla J|$ around the critical curve,
the faster decreases the magnification value and smaller the ``width" of the 
caustic curve. 

If the mass ratio is very small as in a planetary binary lens, then
\opeqn
 (\nabla J)^{\bf C} = 2 \bar\partial J = - 2 \kappa\bar\partial\bar\kappa
  =  4 \kappa \sum_j {\epsilon_j\over (\bar z - \bar x_j)^3} \ ,     
\cleqn
and on the critical curve of a small mass lens $\epsilon_2 << 1$,
\opeqn
|\nabla J| \approx 4 \left| 1 + {\epsilon_2\over x_2^3} \right| 
    \approx {4\over\sqrt{\epsilon_2}}  >> 1 \ .
\cleqn
For terrestrial planets, the mass ratio to the host star is 
$\sim 10^{-5}-10^{-6}$, and the second or high order terms can be dominant 
whether the planet is bound or free-floating. Thus, power expansion 
approximation can not be used for integrating over the stellar luminosity 
profile.  In a multiple
point lens system, the range of $|\nabla J|$ can be very large depending
on the caustic loops. It is safe to say that the central caustic will have
$|\nabla J| \sim {\cal O}(1)$.  In the practice of light curve fitting,
model light curves are all calculated numerically. At caustic crossings,
what may be referred to as ``local ray-shooting method" is used to be
able to accomodate the rapid change of the inverse Jacobian determinant
$J^{-1}$ and instability related to the criticality.

\section*{-2. Conclusion and Discussion}

We have reexamined the quadratic behavior of the lens equation near the critical
curve. In the second order approximation, the lens equation becomes a real 
symmetric quadratic equation between 
the source line and image line locally defined at a proper causic point and 
critical point respectively. In other words, given a source position $\omega$, 
there exits a {\bf preferred caustic point} $\omega_\circ$ such that 
$\delta\omega = \omega - \omega_\circ$ is in the same direction as 
$\bar\partial J (z_\circ)$ where $z_\circ \mapsto \omega_\circ$ under the
lens equation. The source $\omega$ generates two images at $z_\pm$ along
the critical direction defined at $z_\circ$, and the magnification of the 
images is given by $\sqrt{(4\delta\omega_- J_-)^{-1}}$. Source line refers
to $\omega - \omega_\circ$ and image line refers to the line connecting 
$z_\pm$ bisected by $z_\circ$. We emphasized that the image line 
($\pm E_- (z_\circ)$) and source line ($\bar\partial J (\omega_\circ)$)
are not orthogonal to each other. 
In fact, the second order approximation fails when they
are orthogonal because they are orthogonal at cusps ($J_- = \nabla J\cdot E_- =0$).

It is not necessarily a trivial matter to find the {\bf preferred caustic point}
$\omega_\circ$ for each source position $\omega$, say, of a time sequence of
a stellar disk that crosses the caustic curve. It is in fact unncessary 
because we can recover the image positions and their magnifications by power
expanding the lens equation at a nearby but arbitrary caustic point. Fig. 1 
clearly indicates that we have freedom to choose a caustic point which is 
in the neighborhood of the caustic point that defines the (shortest) distance. 

The criticality (vanishing derivative) occurs only in one direction, and the
power expansion of the lens equation at a {\it non-preferred caustic point}
(meaning $\not= \omega_\circ$) involves non-vanishing non-critical component
of $dz$ ({\it i.e.,} $dz_+ \not= 0$) and this non-critical component is 
a linear equation of the source position variable components $\omega_+$ and
$\delta\omega_-$ in the lowest approximation. The non-critical component
$dz_+$ causes a migration along the critical curve and we can see in Fig.3 
that the non-critical component $ 2dz_+ $ measures the shift 
of the non-preferred caustic point from the preferred caustic point in 
the direction of $E_+$.  If we include the nonlinear effect, the shift 
deviates from $ 2dz_+ $.   

In this Taylor expansion approximation, the gradient of the Jacobian determinant
($\nabla J = 2 \bar\partial J$ with the understanding that we freely exchange 
between complex notations and real countparts for the same 2-d vectors) 
is a constant vector calculated at the critical point where the power expansion 
is made (or at the origin). 
And, the directional difference of $\bar\partial J$ at two different
caustic points ($\omega_c$ and $\omega_{c_0}$) can not be incorporated in the 
second order approximation. In the lowest approximation, $dz_+$ is a 
linear equation of $\delta\omega_+$ and $\omega_-$, and so the direction of 
the source line of $\omega$ is given by the source line of a source position 
whose preferred caustic point would be the origin ($\omega_c$). The preferred
caustic point for $\omega$ in this approximation is given by what we denoted
$\omega_{c_0})$ (and marked as $w0$ in Fig. \ref{fig-approx}).  
We emphasize that the preferred caustic point $\omega_{c_0})$ is distinguishable
from what we referred to as the approximate preferred caustic point that is 
determined as the cross section of the linear equation $bx-ay=abu$ and
the quadratic caustic equation. The linear approximation for $dz_+$ is valid
Where the difference is sufficiently small.  

\begin{figure}
\plotone{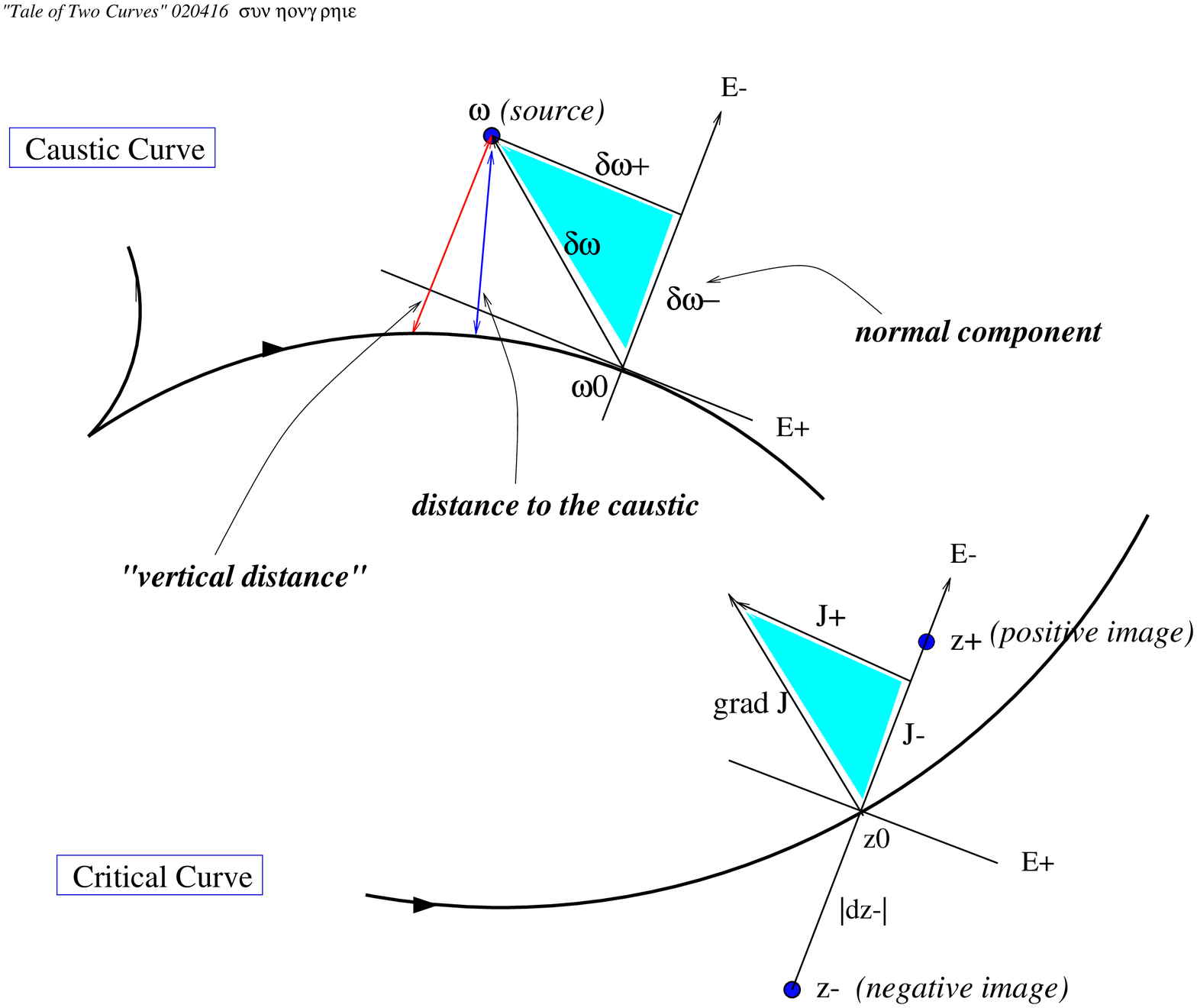}
\figcaption{\label{fig-imageline}
Quadratic behavior of the lens equation near a critical point $z_\circ$ where
$J_- (z_\circ) \not \approx 0$: ~The criticality defines the image line 
~$dz \parallel \pm E_- (z_\circ)$ and 
the source line ~$\delta\omega \parallel \bar\partial J (z_\circ)$, and
the lens equation becomes a real symmetric quadratic equation from the image line 
to the source line: ~$\delta\omega = {1\over 2} \bar\partial J~ dz_-^2$. 
There are two images when $\delta\omega / \bar\partial J > 0$, and no images
when $\delta\omega / \bar\partial J < 0$. 
Here ~$\delta\omega \parallel \bar\partial J$, and $\omega$ generates two 
images $z_\pm$ near the critical point $z_0$. The positive image at $z_+$ is in 
the direction of $\nabla J$, and the negative image at $z_-$ is in the opposite
direction.  
{\it The amplification of the images is $\sqrt{(4\delta\omega_- J_-)^{-1}}$, 
and both the shortest distance and the ``vertical distance" differ from the 
``truely relevant distance" $|\delta\omega_- (\omega_\circ)|$.} 
}
\end{figure}

Once we know the preferred caustic point, the gradient of the 
Jacobian determinant, and the normal to the caustic where both of the latter 
two are calcualted at the origin ($\omega_c$), then the image positions and their
magnifications can be calculated. The magnifications formula can be written in
the same form as those calculated at the preferred caustic point $\omega_\circ$.  
\opeqn
 J = \pm \sqrt {J_- \Delta\omega_-}
\label{eqJDelta}
\cleqn  
$\Delta\omega_-$ is the normal component of the shift of $\omega$ from the 
preferred caustic point $\omega_{c_0}$.  We illustrated in Fig. \ref{fig-approx}
that {\it normal component} and {\it ``vertical distance"}\footnote{
According to The Random House College Dictionary, {\it vertical} is to be in
the same direction as the axis. Here the axis is taken as the critical 
direction vector at the origin ($ E_- (\omega_c)$) where the power series 
expansion is made. The critical vector at the origin is normal to the caustic
curve at the origin.}
are clearly distinguishable both conceptually and practically.
hen the curvature of the caustic can be neglected either because the stellar
disk is sufficiently small or $J_+ \approx 0$ in the neighborhood of the
caustic crossing point, both the ``vertical distance" and the (shortest)
distance become comparable to the normal component.
When $\omega_{c_0}$ and $\omega_c$ coincide with $\omega_\circ$, 
$\Delta\omega_- = \delta\omega_-$, and this symmetric quadratic case is 
depicted in Fig. \ref{fig-imageline}. 
\opeqn
 J = \pm \sqrt {J_- \delta\omega_-}
\label{eqJdelta}
\cleqn

\citet{vertical} raised an issue of the identity of the ``relevant distance".
As we have emphasized many times, the ``truly relevant distance" is 
$ |\Delta \omega_-|$ given as follows (rb99.16 but with the proper substitution
of the linear approximation of $dz_+$ in terms of $\delta\omega_+$ and 
$\delta\omega_-$). 
\opeqn
 \Delta\omega_- \equiv \delta\omega_- - \delta\omega_- (J=0) 
  = \delta\omega_- + {|\partial J|^2 \over J_-} dz_+^2  \ , 
\cleqn   
where the linear function is given as follows.
\opeqn
  (J_- \delta\omega_+ - J_+ \delta\omega_- ) = 2 J_- dz_+ 
\cleqn
We reiterate that $\delta\omega_-$ is the $E_-$ component of 
$\delta\omega = \omega - \omega_c $ where $\omega_c$ is the origin, and 
$\delta\omega_- (J=0)$ is the $E_-$ component of the position vector of
the caustic point ($\omega_{c_0}$) which lies on the source line. 
The criticality dictates two normals, and the lens equation in the second
order approximation is a quadratic relation between the variables along these
two normals. And, the two normals are oblique to each other. 

It should be worth pondering if the notation $\delta\omega_- (J=0)$ is
potentially confusing. What else can 
one imagine for the quantity $\delta\omega (J=0)$ aside from the intrinsically 
relevant quantity that is the derived preferred caustic point position vector
$\omega_{c_0}$? One can drop a straight line in the direction of the 
``vertical axis" from $\omega$ in Fig. \ref{fig-approx}
and claim that the intersection point 
with the caustic curve must be $\delta\omega (J=0)$ and its $E_-$ component be 
$\delta\omega_- (J=0)$. That may seem all right at a first glance.
But, what would be the significance of the line that is parallel to 
$E_-(\omega_c)$ and passes through $\omega$?  We do not know. However, we know 
that the corresponding equation $\delta\omega_+ = constant$ (given by the value 
of the $E_+$ component of the particular source position) can not be derived 
from the quadratic lens equation (\ref{eqxy}). Given the irrelevance, we can 
discard the case from the list of potentially confusing interpretations. 
What else can one imagine for $\delta\omega (J=0)$? Currently, we lack 
imagination for other possibilities of confusion. We take it as a good enough 
reason to pardon our notation and close the case. That is, of course, until 
someone brings a brilliant confusion candidate to our attention. 
 
\begin{enumerate}
\item
The interpretation of \citet{vertical} that $ |\Delta \omega_-|$ is the
``vertical distance" is an erroneous misinterpretation.
\item
\citet{vertical} claims to have reproduced the results in SEF.  
We find that the second equation of SEF.6.17 has a mistake. By now, we have
a clear understanding that when we expand the lens equation in the neighborhood
of an arbitrary non-preferred caustic point, there will be terms 
describing a migration along the critical curve. The condition
for migration is $0 = dJ = a u + bv$, and so 
we expect $v = - au/b + ...$. If the two equations in SEF.6.17 are compared
($x_1$ is correct), it should be clear that the missing term is $a^2 y/( 2 b^2)$
in our notation. 

\begin{enumerate}
\item
The fault is at the truncation prescription in the first paragraph in SEF.p.189.
If we look at the RHS of equation (12), the most dominant term is $v^2$ term, 
and so $v \approx \sqrt{y}$. We know that that is exactly the case, namely
$v \propto \sqrt{y}$, when the origin is the preferred caustic point and so
there is no migration along the critical curve. If we look at equation (11),
it is clear that the dominant term in the RHS is $v^2$ term and so $u$ is 
linear in $x$ and $y$ as is shown in equation (14). Thus, $v^2 \sim x \sim y \sim u$,
and they are sufficiently smaller than 1 (Einstein ring radius). Incidentally, 
that is why $v$ is much larger than the other three, and images across the 
critical curve are tangentially stretched large.  Since the migration condition
is linear in $u$ and $v$, we expect $v$ to be made of terms of two different orders:
$\sqrt{y}$ and linear terms in $x$ and $y$. Therefore, all the terms up to the
second order in $y$ (so, $xv\sim v^3, xv^2 \sim v^4$) should be kept to calcualte 
$v$ from the second equation of SEF.6.17. The linear terms in $x$ and $y$ 
come from $xv$ and $v^3$ terms respectively.

\item
One may suggest to ignore the linear term in $y$ because that is smaller than
$\sqrt{y}$ term. One problem is that the linear term in $y$ does not have any
reason to be small in comparison to the linear term in $x$, which is reflected
in the first equation in SEF.6.17. In other words, there is no reason to 
truncate $v^3$ term while keeping $xv$ because they are of the same order. 
Thus, there is no foundation for the prescription
to ignore the linear term in $y$ while keeping the linear term in $x$. In fact, 
dropping the linear term in $y$ causes a conceptual inconsistency by obscuring 
the geometric nature of the linear terms -- migration along the critical curve.    
One may argue that $a = J_+ \approx 0$, of course. Then $\Delta\omega_-$ approches
$\delta\omega_-$, and the (shortest) distance and ``vertical distance" converge
to the normal component. In general, we should keep the both linear terms.
\end{enumerate}
The missing term $a^2 y/( 2 b^2)$ affects the expression of $\Delta\omega_-$.
If one cosmetically interprets the erroneous expression, one can be led to
the erroneous conclusion on the distance drawn by \citet{vertical}.
The fault may lie in that the apparent relevance of the {\it vertical distance} 
which may have been a pleasing discovery for a brief moment has not been
tested through the routine logical digestion process necessary for a meaningful
claim and has been put forward as a fact. The correct relevant distance
is the {\it normal component} $\delta\omega_- - \delta\omega_- (J=0)$
where the normal direction is given by the critical direction at the origin 
where the functional definition of the origin derives from that that is where
the Taylor (or power) expansion is made.

\item
The error in rb99 we described somewhat extensively in a previous section 
is propagated to rb99.13 through rb99.16. The substitution 
$\delta\omega_+ \Leftarrow 2 dz_+$ 
in those equations should be replaced by the correct substitution
$\delta\omega_+ - \delta\omega_- J_+/J_-  \Leftarrow 2 dz_+$.

\end{enumerate}

\section*{ \ \ \bf  -3. Epilogue: Another Long Walk to Renaissance }

\citet{vertical} emphasized to have worked out in (2-d) real variables.  
They made an erroneous conclusion and kindly warned the readers accordingly, 
{\it `there are not many people who know
that the distance is the ``vertical distance".'} 
This claim may imply a correlation between the error and the choice of 
variables, but that is not the case. 
The missing term in the second equation of SEF.6.17 could have
been recovered if one had stared at the two equations of SEF.6.17 and scratched   
the head about the meanings of the linear terms in comparison to the square 
root terms. In our humble belief, (astro)physically meaningful quantities
have intrinsic values (many times as the form of geometric quantities) and
offer intuitive understanding.  

One advantage of the complex variable is that quasi-analytic lens equations  
are, well, quasi-analytic. The derivatives are completely determined by one 
analytic function. There is nothing much great about it except that it is 
simpler. 
The system is constrained and the constraint is completely contained within the 
nature of the variable, and so it is easier to get to know the system. 
Especially, for one variable analytic function, we even know how to integrate
because the first theorem of analysis is simply defined. 

Is there a taboo against complex variables in astro-physics? If there were, 
it must be the time to change now that astro-physicists are gearing up to  
the idea of testing red-shift dependence of the electromagnetic fine 
structure constant -- the beloved prime number 137 when inversed and truncated.    
(S. Bechwith, HSL workshop, U.Chicago, April, 2002; 
See J.K.Webb et al., PRL, 87, 091301 (2001) = astro-ph/0101375 
for a $\sim 2\sigma$ level claim of 
$\Delta\alpha/\alpha \sim - 10^{-5}$ from quasar absorption 
spectrum analyses\footnote{We do not pretend to have understood the systematics
related to the analyses. Some useful numbers can be found in Table 1 of 
M.T.Murphy et al., astro-ph/0012421 and they indicate that the usage of the
rest frame UV lines requires a precision level of $\sim 10^{-8}$ in the spectral
measurements to be able to test $\Delta\alpha/\alpha \sim - 10^{-5}$. 
The resolution of $\sim 7 km/sec$ obtained from HIRES/Keck I 
is far worse than the necessary precision, and the analyses rely on 
broad spectral profile fitting and $\chi^2$ estimation.}; 
many more papers in the usual suspect sites astro-ph and hep-ph; 
P.A.M.Dirac may have been the first to be serious enough to write a paper,
Nature, 139, 323 (1937), on the variability of the fundamental constants 
including the fine structure constant; 
the first Kaluza-Klein models were written in the mid-1930's as well; 
see T. Chiba, gr-qc/0110118 for a review of constraints on the variation of 
fundamental constants; coupling constants are believed to run with energy; 
there are strong experimental indications for grand unification of the electroweak 
and strong interactions (see F. Wilczek, hep-ph/0101187 for a summary)  
and relevance of supersymmetry\footnote{We happen to be a believer of 
supersymmetry because that seems to uniquely naturally solve the 
``missing spin 3/2 problem".
No fundamental spin 3/2 particles have been seen yet. Of course, Higgs particles
have not been found either. Once named God particles, they may be shy away from
the physical nature. LHC will tell us within its limit, no doubt.} .) 
 
Gravitational lensing is an old subject revitalized with sensitive and gigantic
instruments. Leonardo da Vinci seems to stand out as a Renaissance man. 
And, Mona Lisa a Renaissance woman. 
Mona Lisa is believed to be da Vinci himself sanc frizzy hair and beard. 
The mysterious smile of Mona Lisa may be a smirk of da
Vinci daring the admirers, ``You are looking at me." 
Or, ``Can we live without phases?" It is the time to release complex variables 
from the hit list and allow some freedom. In fact, 
eigenvectors, eigenvalues, and all that are about the differential behavior
in the underlying 2-d space. $E_\pm$ decomposition coefficients are all real.    
It is inevitable to deal with two real variables in one way or another. 
Then, what significance does it carry to claim to have chosen real
basis vectors (implitly as two column vectors $(1, 0)$ and $(0, 1)$) instead of 
$E_\pm$? It is unclear. The authors \citet{vertical} did not specify  
the qualification of the statement for one purpose or another. 
It remains to be a mysterious invitation to a dark age.


\def\ref@jnl#1{{\rm#1}}
\def\aj{\ref@jnl{AJ}}
\def\apj{\ref@jnl{ApJ}}
\def\apjl{\ref@jnl{ApJ}}
\def\apjs{\ref@jnl{ApJS}}
\def\aap{\ref@jnl{A\&A}}
\def\aapr{\ref@jnl{A\&A~Rev.}}
\def\aaps{\ref@jnl{A\&AS}}
\def\mnras{\ref@jnl{MNRAS}}
\def\prl{\ref@jnl{Phys.~Rev.~Lett.}}
\def\pasp{\ref@jnl{PASP}}
\def\nat{\ref@jnl{Nature}}
\def\iauc{\ref@jnl{IAU~Circ.}}
\def\aplett{\ref@jnl{Astrophys.~Lett.}}
\def\annrev{\ref@jnl{Ann.~Rev.~Astron.~and Astroph.}}

\clearpage


%
%

\begin{figure}
\plotone{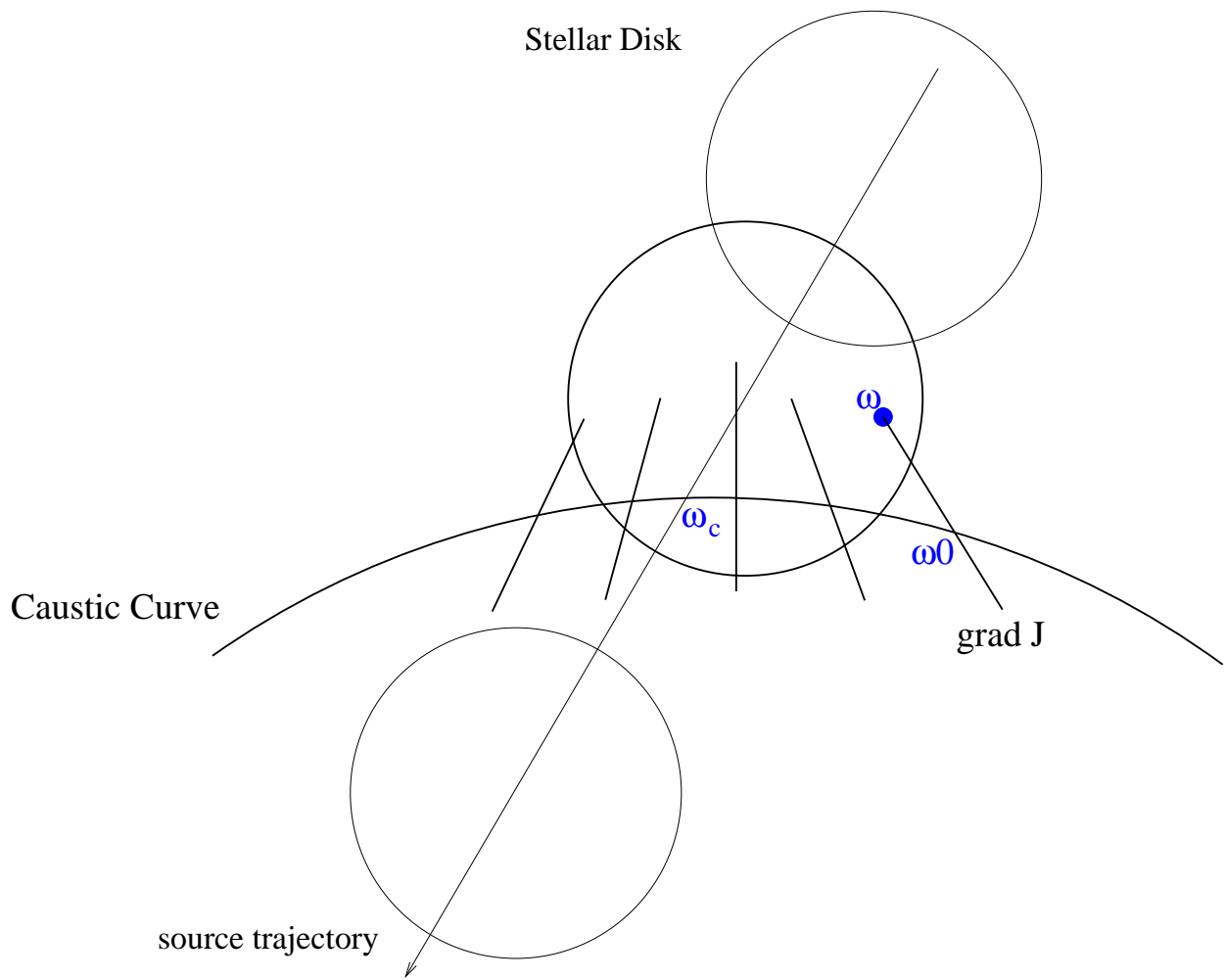}
\figcaption{\label{fig-gradj}
If we consider a large stellar disk, the source line ($\bar\partial J$) for each
point in the disk can be diverse. Given a source position $\omega$ inside the 
caustic curve, $\omega_0$ will define the source line, and the image line of 
the two images will be at $z_\circ$: $z_\circ \mapsto \omega_0$ as we have discussed
in the text and the caption of Fig.1. One issue is how well we can recover
the images $z_\pm$ and their amplification $\sqrt{(4\delta\omega_- J_-)^{-1}}$
when we take a nearby but arbitrary caustic point $\omega_c$ as the origin
where the coordinate system $\{ E_+, E_- \}$ is defined and the power
series expansion coefficients are calculated.
}
\end{figure}

\begin{figure}
\plotone{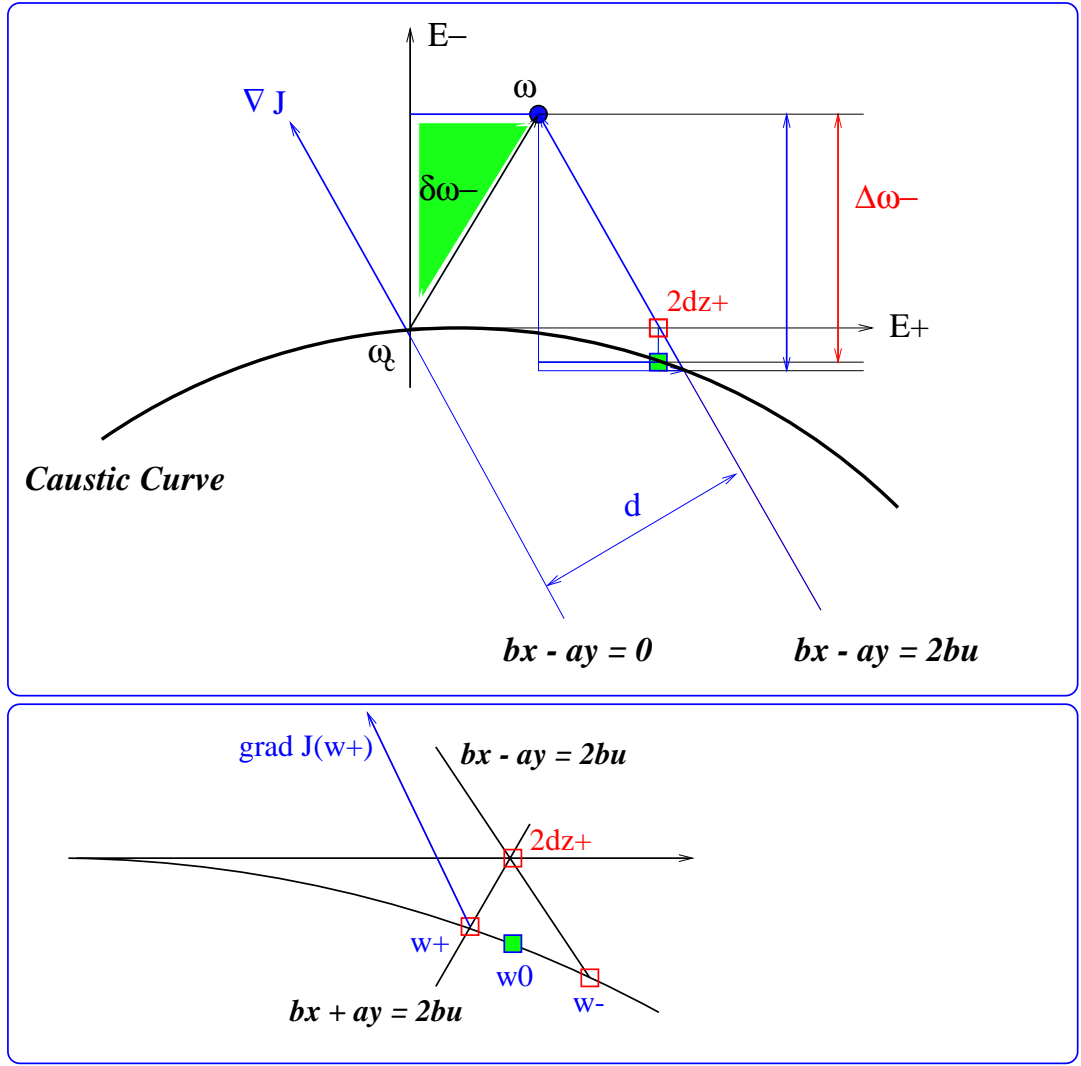}
\figcaption{\label{fig-approx}
Non-preferred caustic point $\omega_c$ as the origin. 
The value of multiplication function $J_-$ is calculated at the origin, and
$\Delta\omega_-$ is the normal component of $\delta\omega - \delta\omega (J=0)$. 
The bottom panel compares the caustic points: $w+$ is the correct caustic point 
for given $2 dz_+$;  $w-$ is the caustic point defined by the line 
$bx - ay = 2bu$ and was referred to as the approximate preferred caustic point
for $\omega$; $w0$ is the derived preferred caustic point for $\omega$ in the
linear approximation of $dz_+$.} 
\end{figure}

\begin{figure}
\plotone{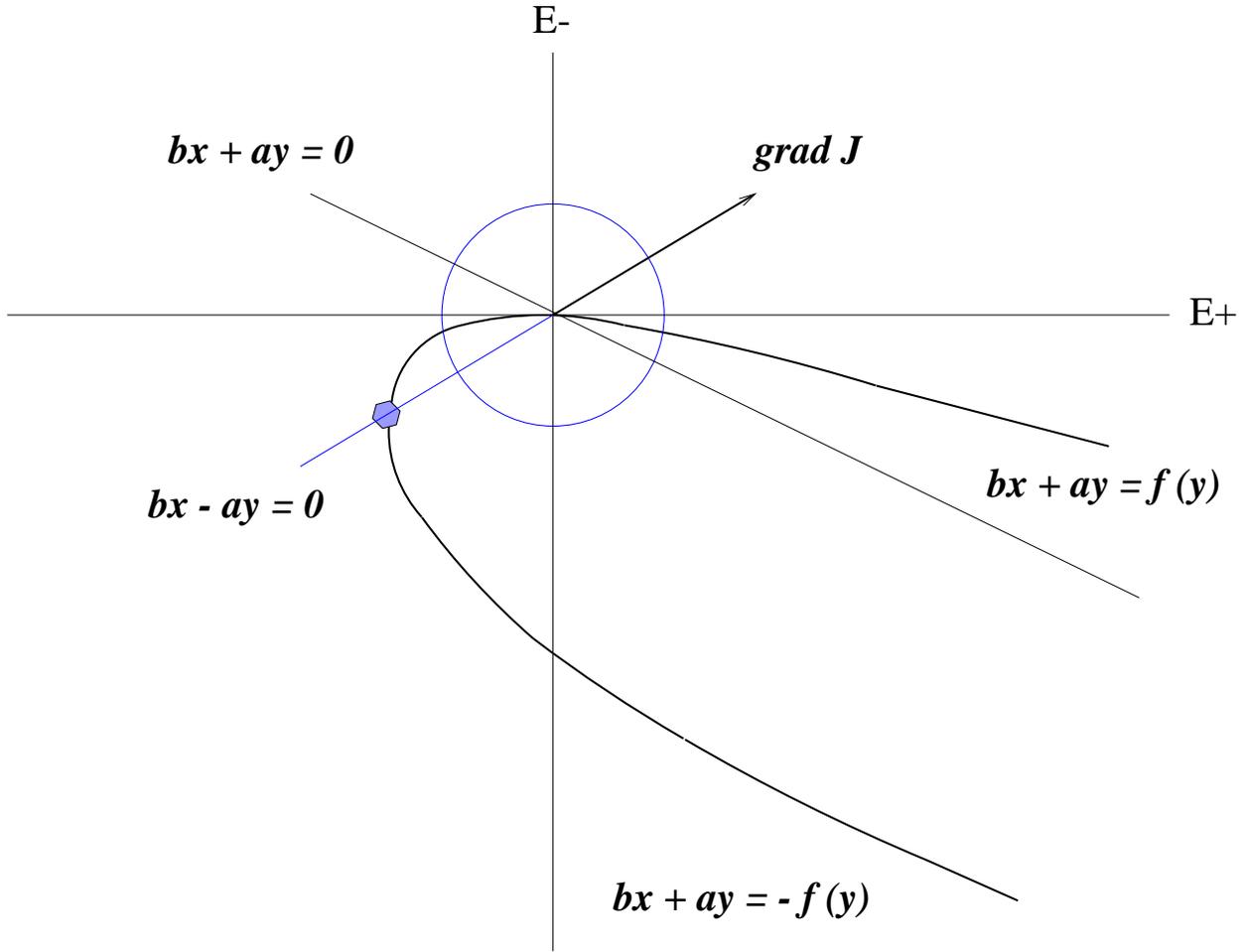}
\figcaption{\label{fig-caustic}
The caustic curve in the second order approximation satisfies a quadratic equation  
between $y$ and $bx+ay$. Here $f(y) = \sqrt{-16 y b^3 /(\nabla J)^2}$. The 
equation is a valid description of the original caustic curve in the neighborhood
(here circled for illustration) where $|ay|$ is negligible in comparison to the
other two terms $bx$ and $f(y)$ and becomes a quadratic function between $y$ 
and $x$. 
}
\end{figure}


\begin{thebibliography}{}

\bibitem[\protect\citeauthoryear{{Chang}}{{Chang}
  }{1984}]{ch84}
{Chang}, K. 1984, A \& A 130, 157 

\bibitem[\protect\citeauthoryear{{Chang} \& Refsdal}{{Chang} \& Refsdal 
  }{1979}]{cr79}
{Chang}, K., and Refsdal, S. 1979, \nat, 282, 561 

\bibitem[\protect\citeauthoryear{{Gaudi} \& Petters}{{Gaudi} \& Petters 
  }{2001}]{vertical}
{Gaudi}, S., and Petters, A. 2001, astro-ph/0112531 

\bibitem[\protect\citeauthoryear{{Rhie}}{{Rhie}
  }{1997}]{rh97}
{Rhie}, S. H.  1997, \apj, 484, 63

\bibitem[\protect\citeauthoryear{{Rhie}}{{Rhie}
  }{2001}]{quadlens}
{Rhie}, S. H.  2001, astro-ph/0103463 

\bibitem[\protect\citeauthoryear{{Rhie} \& Bennett}{{Rhie} \& Bennett
  }{2001}]{binary}
{Rhie}, S.~H., and Bennett, D.~P. 2001, ``Notes on Gravitational Binary
Lenses" (unpublished) 

\bibitem[\protect\citeauthoryear{{Rhie} \& Bennett}{{Rhie} \& Bennett 
  }{1999}]{limb} 
{Rhie}, S.~H., and Bennett, D.~P. 1999 (rb99), astro-ph/9912050 

\bibitem[\protect\citeauthoryear{{Schneider}, Ehlers, \& Falco}{{Schneider}, 
  Ehlers, \& Falco}{1992}]{sef} 
{Schneider}, P., Elhers, J., and Falco, E. 1992 (SEF), {\it ``Gravitational Lenses"},
 Springer-Verlag  



\end{thebibliography}
\end{document}